\documentclass[]{article}
\title {Four New Single-Error-Correcting Binary Codes}  
\author{Moshe Milshtein \\
	\small Sarasota, Florida, USA \\
	\small \texttt{moshe.milshtein@gmail.com}
}

\date{}

\usepackage{booktabs}
\usepackage{makecell}
\usepackage{amsmath}
\usepackage{amssymb}
\usepackage{url}
\usepackage[numbers]{natbib}

\begin{document}

\maketitle

\begin{abstract}
We construct a \((16, 3072, 3)\)-code, a \((17,6144,3)\)-code,  an \((18,12288,3)\)-code and a \((19,24576,3)\)-code. Consequently, we get four new lower bounds: \(A(16,3) \geq 3072\), \(A(17,3) \geq 6144\), \mbox{\(A(18,3)  \geq 12288\)} and \(A(19,3) \geq~24576\). Each of the  codes consists of a union of cosets of a linear code. The constructions are the result of a computer search executing  a tabu search algorithm.
\end{abstract}

\section{Introduction}

A fundamental problem in Coding Theory is  to find a \(q\)-ary block code of codeword length \(n\) and minimum Hamming distance \(d\) with the maximum possible number of codewords, for given integers \(q\), \(n\) and \(d\). This optimal number of codewords is denoted by \(A_q(n,d)\), and for \(q=2\) simply by \(A(n,d)\). An  \(M\)-codeword  binary code  of length \(n\) with minimum Hamming distance \(d\)  is called an \((n,M,d)\)-code.

For small values of~\(n\), Brouwer~\cite{brouwertable} maintains an online table of the best known lower and upper bounds on \(A(n,d)\). For \(n \leq 15\) the exact values of \(A(n,d)\) are known, but for larger values of \(n\), in most cases only  lower and upper bounds were determined. 

In this paper we construct four new binary codes:  a \((16, 3072, 3)\)-code, a \((17,6144,3)\)-code,  an \((18,12288,3)\)-code and a \((19,24576,3)\)-code. The existence of the codes implies four new lower bounds on \(A(n,d)\) as detailed in Table~\ref{tab:bounds}. We found the codes using a computer  search  for  a suitable union of cosets of a linear code.

\begin{table}
	\centering
	\begin{tabular}{rrrrrl}
		\toprule
		\(n\) & \(d\) &  \makecell{Known\\Lower Bound} & \makecell{Improved\\Lower Bound} & \makecell{Known\\Upper Bound}  & References\\
		\midrule
		\(16\) & \(3\) &  \(2816\) &  \(\mathbf{3072}\) &  \(3276\) & \cite{MILSHTEIN2015975, LAAKSONEN201765, best1978bounds} \\
		\(17\) &  \(3\) &  \(5632\) &  \(\mathbf{6144}\)&  \(6551\) & \cite{LAAKSONEN201765, tseng_semidefinite_2023} \\
		\(18\) &  \(3\) &  \(10496 \)&  \(\mathbf{12288}\) &  \(13104\)  &  \cite{Hamalainen,best1978bounds}  \\
		\(19\) &  \(3\) &  \(20480 \)&  \(\mathbf{24576}\) &  \(26168\)  &  \cite{best20480,Haas_2008}  \\
		\bottomrule
	\end{tabular}
	\caption{Improved Lower Bounds on \(A(n,d)\)}
	\label{tab:bounds}
\end{table}

The rest of the paper is organized as follows. Section~\ref{algorithm} describes the search algorithm we used. Section~\ref{experiments} provides  some details about the computational experiments and includes a link to the results.

\section{The Search Algorithm}
\label{algorithm}

The problem of finding the maximum possible size of a \(q\)-ary \((n,M,d)\)-code is a combinatorial optimization problem. One way to make the problem more accessible is to restrict the search to codes that consist of cosets of a linear code. This restriction was used to find some lower bounds on \(A(n,d)\) for small values of \(n\) \cite{ostergard_two_2005, elsell}.  We follow this method here. We use stochastic local search rather than complete search for finding a set of suitable cosets. More specifically, we use the tabu search metaheuristic \cite{glover1989tabu}.

Let  \(C\) be a linear code of length \(n\), dimension \(k\) and minimum distance at least \(d\). We define a graph \(G\)  whose vertices correspond to the cosets of the linear code \(C\), and whose edges connect two vertices if the corresponding cosets are at distance less than \(d\) from each other.  Let \(U\) be  an \(m\)-vertex independent set in the graph.  The union of the cosets that correspond to the vertices of \(U\) is an \(m 2^k\)-codeword code of length \(n\) with minimum distance at least \(d\). Therefore,  finding  an \(m\)-vertex independent set  in the graph \(G\) yields a good code if \(m\) is large enough.

For solving the independent set existence  problem we  use a stochastic local search algorithm. The  general outline of a stochastic local search algorithm is as follows \cite{HOOS200561}. We start from an initial candidate solution. At every iteration we move from the current candidate solution \(s\) to another candidate solution \(s'\). The new candidate solution \(s'\) is chosen from a set of candidate  solutions called the neighborhood of \(s\), using some stochastic criteria. An evaluation function that assigns a value to every candidate solution guides the choice of the neighbor \(s'\). A candidate solution that meets the problem requirements is called a solution, and if we encounter one the algorithm is successful.

For the rest of this section we describe in detail the search algorithm we used.

The input for our algorithm consists of integers \(n\) (the codeword length), \(d\) (a minimum distance), \(m\) (the number of cosets in the union we are trying to find),  \(k\) (the dimension of the linear code), and  a linear code \(C\) of length \(n\), dimension \(k\) and minimum distance at least \(d\). Additionally, we have two tabu search parameters: \texttt{tt} (tabu tenure) and \texttt{maxNoRecordIter} (the allowed maximum  number of consecutive  iterations that do not improve the best  evaluation value seen so far).

Let \(G\) be the graph with vertices corresponding to the cosets of \(C\)  and edges connecting cosets with distance less than \(d\) from each other.  The output is either an \(m\)-vertex independent set  in \(G\) or  \verb|FAIL| if the algorithm cannot find such a set.

We define a \emph{candidate solution} as  an \(m\)-vertex subset of the graph \(G\). A \emph{solution} is  a candidate solution that is an independent set. We define a \emph{neighbor} of a candidate solution \(s\) as a candidate solution obtained by replacing a vertex from \(s\) with a vertex that is not in \(s\) (a \emph{swap} move).

We now define an \emph{evaluation function} that assigns to each candidate solution \(s\) a value that reflects in some sense the quality \(s\). Larger evaluation values suggest worse quality of the \(m\)-subset. 

First, let \(\mathbf{u}\) and \(\mathbf{v}\) be vectors in \(\mathbb{F}_2^n
\), and let \(d_H(\mathbf{u}, \mathbf{v})\) be the Hamming distance between them. We define a cost function
\[
\operatorname{cost}(\mathbf{u}, \mathbf{v}) = 
\begin{cases}
	0 & \text{if } d_H(\mathbf{u}, \mathbf{v}) \geq d, \\
	(d - d_H(\mathbf{u}, \mathbf{v}))^2 & \text{otherwise.}
\end{cases}
\]

Next, we assign weights to the edges of \(G\).  Let \(e\) be an edge connecting two distinct cosets \(C_1\), \(C_2\) of the linear code \(C\). The weight of the edge  \(e\) is defined naturally  by
\begin{equation*} 
	\operatorname{weight}(e) = \sum_{\mathbf{c_1} \in C_1, \mathbf{c_2} \in C_2} \operatorname{cost}(\mathbf{c_1}, \mathbf{c_2} ).
\end{equation*}

Finally, the evaluation value of a candidate solution \(s\) is the  sum of the edge weights in the subgraph induced by \(s\). The evaluation value is \(0\) if and only if \(s\) is a solution.

The \emph{initial candidate solution} is  an \(m\)-vertex subset chosen uniformly at random. A swap is declared  \emph{tabu} if it involves  inserting a vertex that was removed in the last \texttt{tt} iterations. The \emph{aspiration criterion}  is met by any move that leads to a neighbor with an evaluation value that is smaller than any evaluation value encountered in previous iterations. An \emph{admissible move} is a move that is not tabu or meet the aspiration criterion.  At any iteration our next move is  a uniformly random best admissible move, that is, we only consider an admissible move if no other admissible move leads to a lower valued neighbor.

The algorithm stops if one of the following occurs: a solution is found, the record evaluation value did not improve  in the last  \texttt{maxNoRecordIter} iterations or  no admissible moves exist.

\section{Computational Experiments and Results}
\label{experiments}

We refer to the algorithm described in Section~\ref{algorithm} as the \emph{basic algorithm}. We now describe  the computational experiments we performed with the basic algorithm.

We picked some \(n\) and \(d\) for which we want to improve the best known lower bound \(l \leq A(n,d)\). We set \mbox{\(m = \lceil (l  + 1) / 2^k \rceil \)} to try the minimum possible improvement.  We chose some \(k\), and generated all the inequivalent linear codes of length \(n\), dimension \(k\) and minimum distance at least \(d\). For the linear codes classification we used the \verb|Generation| program from the software project \verb|QextNewEdition|~\cite{bouyuklievQextNewEdition, classification}. For each linear code we executed the basic algorithm with increasing tabu tenures \(2^0,2^1, 2^2,\ldots\), one execution per tabu tenure.    If the tabu tenure became large enough so that no admissible moves existed, or if we reached tabu tenure with value \(2^{20}\) we moved to the next linear code. We usually set the \texttt{maxNoRecordIter} parameter to \(100{,}000\). 

In Table~\ref{tab:solution_parms} we give the combinations of the parameters \(n\), \(d\) and \(k\)  that yielded a solution. The size of the first solution we found appears in the column with the header \emph{Initial \(m\)}. This solution already implies  a record-breaking code. In order to find a larger solution we executed a mix of the following procedures:

\begin{itemize}
	\item We added to the solution a vertex that was not adjacent to any  solution vertex. This was possible quite often.
	\item We kept the successful linear code and tabu tenure that yielded the solution fixed, increased \(m\) somewhat and repeated the basic algorithm up to \(1{,}000\) times.
	\item We increased \(m\) and  repeated the search over all the inequivalent linear codes of length \(n\), dimension \(k\) and minimum distance at least \(d\). 
	
\end{itemize}

The size of the largest solution we found after experimenting with the above three procedures appears in the column with the header \emph{Largest \(m\)}.

\begin{table}
	
	\centering
	\begin{tabular}{rrrrr}
		\toprule
		\(n\) & \(d\) & \(k\) & Initial \(m\) & Largest \(m\) \\
		\midrule
		\(16\) & \(3\) &  \(4\) &  \(177\) & \(192\)  \\
		\(17\) &  \(3\) &  \(4\) &  \(353\)&  \(384\)   \\
		\(18\) &  \(3\) &  \(5 \)&  \(329\) & \(384\)  \\ 
		\(19\) &  \(3\) &  \(5 \)&  \(641\) & \(768\) \\ 
		\bottomrule
	\end{tabular}
	\caption{Parameters of The Successful Runs} 
	\label{tab:solution_parms}
\end{table}

The total wall-clock run time of our program was several weeks, including time spent on failed attempts to further improve the lower bounds given in Table~\ref{tab:bounds}  and time spent trying to improve other \(A(n,d)\) lower bounds. The program   ran on two computers: one with an AMD Ryzen 9 5950X 16-core processor and the other with  an Intel Core i7-14700K 20-core processor. The implementation language was C++.

The codes we found can be accessed in \cite{milshtein_2026_21633500}. For each code we provide two files that describe the construction exactly as it was found in the search. One file contains the generator matrix of the linear code and the other contains the coset leaders. 

The bounds  \(A(16,3) \geq 3072\), \(A(17,3) \geq 6144\),  \(A(18,3)  \geq 12288\) and  \(A(19,3) \geq~24576\) follow from the four new codes. The first three bounds also immediately follow from the  bound  \(A(19,3) \geq~24576\) since \mbox{\(A(n-1,d) \geq {A(n, d)}   / 2\).}

\bibliographystyle{plainnat}
\bibliography{MilshteinFourNewCodesBibliography}

@misc{brouwertable,
	title={Table of general binary codes},
	author={Brouwer, Andries},
	url={https://www.win.tue.nl/aeb/codes/binary-1.html},
	year={2018},
	urldate={2026-07-05}
}

@ARTICLE{best1978bounds,
	author={Best, M. and Brouwer, A. and MacWilliams, F. and Odlyzko, A. and Sloane, N.},
	journal={IEEE Transactions on Information Theory}, 
	title={Bounds for binary codes of length less than 25}, 
	year={1978},
	volume={24},
	number={1},
	pages={81-93},
	doi={10.1109/TIT.1978.1055827}
}

@article{MILSHTEIN2015975,
	title = {A new binary code of length 16 and minimum distance 3},
	journal = {Information Processing Letters},
	volume = {115},
	number = {12},
	pages = {975-976},
	year = {2015},
	issn = {0020-0190},
	doi = {https://doi.org/10.1016/j.ipl.2015.07.001},
	author = {Moshe Milshtein}
}

@article{LAAKSONEN201765,
	title = {Constructing error-correcting binary codes using transitive permutation groups},
	journal = {Discrete Applied Mathematics},
	volume = {233},
	pages = {65-70},
	year = {2017},
	issn = {0166-218X},
	doi = {https://doi.org/10.1016/j.dam.2017.08.022},
	author = {Antti Laaksonen and Patric R.J. Östergård}
}

@article{ostergard_two_2005,
	title = {Two {New} {Four}-{Error}-{Correcting} {Binary} {Codes}},
	volume = {36},
	issn = {1573-7586},
	doi = {10.1007/s10623-004-1723-3},
	number = {3},
	journal = {Designs, Codes and Cryptography},
	author = {Östergård, Patric R. J.},
	month = sep,
	year = {2005},
	pages = {327--329}
}

@ARTICLE{elsell,
	author={Elssel, K. and Zimmermann, K.-H.},
	journal={IEEE Transactions on Information Theory}, 
	title={Two new nonlinear binary codes}, 
	year={2005},
	volume={51},
	number={3},
	pages={1189-1190},
	doi={10.1109/TIT.2004.842767}
}

@ARTICLE{classification,
	author={Bouyukliev, Iliya and Bouyuklieva, Stefka and Kurz, Sascha},
	journal={IEEE Transactions on Information Theory}, 
	title={Computer Classification of Linear Codes}, 
	year={2021},
	volume={67},
	number={12},
	pages={7807-7814},
	doi={10.1109/TIT.2021.3114280}
}

@misc{bouyuklievQextNewEdition,
	author       = {Bouyukliev, Iliya and Bouyuklieva, Stefka},
	title        = {QextNewEdition},
	subtitle     = {A Software Package for Classification of Linear Codes},
	year         = {2019},
	url          =   {https://www.moi.math.bas.bg/moiuser/%7edata/Software/QextNewEdition},
	urldate      = {2026-07-10},
	institution   = {Institute of Mathematics and Informatics, Bulgarian Academy of Sciences},
	note         = {Standard version v1.1 [software]},
	version      = {1.1}
}

@article{tseng_semidefinite_2023,
	title = {Semidefinite programming bounds for binary codes from a split {Terwilliger} algebra},
	volume = {91},
	issn = {1573-7586},
	doi = {10.1007/s10623-023-01250-4},
	number = {10},
	journal = {Designs, Codes and Cryptography},
	author = {Tseng, Pin-Chieh and Lai, Ching-Yi and Yu, Wei-Hsuan},
	month = oct,
	year = {2023},
	pages = {3241--3262},
}

@ARTICLE{Hamalainen,
	author={Hamalainen, H.O.},
	journal={IEEE Transactions on Information Theory}, 
	title={Two new binary codes with minimum distance three}, 
	year={1988},
	volume={34},
	number={4},
	pages={875-},
	doi={10.1109/18.9787}
}

@article{Haas_2008, title={On the Failing Cases of the Johnson Bound for Error-Correcting Codes}, volume={15}, DOI={10.37236/779}, abstractNote={&lt;p&gt;A central problem in coding theory is to determine $A_q(n,2e+1)$, the maximal cardinality of a $q$-ary code of length $n$ correcting up to $e$ errors. When $e$ is fixed and $n$ is large, the best upper bound for $A(n,2e+1)$ (the binary case) is the well-known Johnson bound from 1962. This however simply reduces to the sphere-packing bound if a Steiner system $S(e+1,2e+1,n)$ exists. Despite the fact that no such system is known whenever $e\geq 5$, they possibly exist for a set of values for $n$ with positive density. Therefore in these cases no non-trivial numerical upper bounds for $A(n,2e+1)$ are known.&lt;/p&gt;&lt;p&gt; In this paper the author demonstrates a technique for upper-bounding $A_q(n,2e+1)$, which closes this gap in coding theory. The author extends his earlier work on the system of linear inequalities satisfied by the number of elements of certain codes lying in $k$-dimensional subspaces of the Hamming Space. The method suffices to give the first proof, that the difference between the sphere-packing bound and $A_q(n,2e+1)$ approaches infinity with increasing $n$ whenever $q$ and $e\geq 2$ are fixed. A similar result holds for $K_q(n,R)$, the minimal cardinality of a $q$-ary code of length $n$ and covering radius $R$. Moreover the author presents a new bound for $A(n,3)$ giving for instance $A(19,3)\leq 26168$.&lt;/p&#38;gt;}, number={1}, journal={The Electronic Journal of Combinatorics}, author={Haas, Wolfgang}, year={2008}, month={Apr.}, pages={\#R55} }

@ARTICLE{best20480,
	author={Best, M.},
	journal={IEEE Transactions on Information Theory}, 
	title={Binary codes with a minimum distance of four (Corresp.)}, 
	year={1980},
	volume={26},
	number={6},
	pages={738-742},
	doi={10.1109/TIT.1980.1056269}
}

@article{glover1989tabu,
	author = {Glover, Fred},
	title = {Tabu Search—Part I},
	journal = {ORSA Journal on Computing},
	volume = {1},
	number = {3},
	pages = {190-206},
	year = {1989},
	doi = {10.1287/ijoc.1.3.190}
}

@misc{milshtein_2026_21633500,
	author       = {Milshtein, Moshe},
	title        = {Generator Matrices and Coset Leaders for "Four New
	Single-Error-Correcting Binary Codes"
	},
	month        = jul,
	year         = 2026,
	publisher    = {Zenodo},
	version      = {v1.0.0},
	doi          = {10.5281/zenodo.21633500},
	url =       {https://doi.org/10.5281/zenodo.21633500},
	note         = {Version 1},
}

@incollection{HOOS200561,
	title = {2 - SLS METHODS},
	booktitle = {Stochastic Local Search},
	publisher = {Morgan Kaufmann},
	address = {San Francisco},
	pages = {61-112},
	year = {2005},
	series = {The Morgan Kaufmann Series in Artificial Intelligence},
	isbn = {978-1-55860-872-6},
	doi = {https://doi.org/10.1016/B978-155860872-6/50019-6},
	url = {https://www.sciencedirect.com/science/article/pii/B9781558608726500196},
	author = {Holger H. Hoos and Thomas Stützle}
}

\end{document}